\documentclass[useAMS,usenatbib]{mn2e}
\usepackage{times}
\usepackage{color}
\usepackage[fleqn]{amsmath}
\usepackage{aas_macros,amsfonts,amssymb}
\usepackage{url,hyperref}
\usepackage{graphicx}
\usepackage{booktabs}
\usepackage{siunitx}
\hypersetup{colorlinks=true,linkcolor=black,citecolor=black,filecolor=black,urlcolor=black}
\newcommand{\aCB}{\ensuremath{\alpha\,{\rm Cen}\,{\rm B }}}
\DeclareMathOperator*{\argmax}{arg\,max}
\newcommand{\aCA}{\ensuremath{\alpha\,{\rm Cen}\,{\rm A }}}
\newcommand{\drv}{\ensuremath{\Delta {\rm RV}}}

\newcommand{\lrhk}{\ensuremath{\log R'_{\rm HK}}}
\newcommand{\bis}{\ensuremath{{\rm BIS}}}

\newcommand{\mps}{\ensuremath{\rmn{m}~\rmn{s}^{-1}}}

\title[No planet for Alpha Cen B]{Ghost in the time series: no planet for \mbox{Alpha Cen B}}
\author[V.~Rajpaul et al.]{V.~Rajpaul,$^{1}$\thanks{E-mail: Vinesh.Rajpaul@physics.ox.ac.uk} S.~Aigrain,$^{1}$ and S.~Roberts$^{2}$ 
\\
	$^{1}$Sub-department of Astrophysics, Department of Physics, University of Oxford, Oxford OX1 3RH, UK\\
 	$^{2}$Pattern Recognition and Machine Learning Group, Department of
 	Engineering Science, University of Oxford, Oxford OX1 3PJ, UK\\
 	}
\begin{document}
\date{Accepted 2015 October 19. Received 2015 October 18; in original form 2015 September 27}
\pagerange{\pageref{firstpage}--\pageref{lastpage}} \pubyear{2015}
\maketitle
\label{firstpage}
%------------------------------------------------------------------------------------------------------------------------------
\begin{abstract}
%------------------------------------------------------------------------------------------------------------------------------
We re-analyse the publicly available radial velocity (RV) measurements for Alpha Cen B, a star hosting an Earth-mass planet candidate, Alpha Cen Bb, with $3.24$~d orbital period.  We demonstrate that the $3.24$~d signal observed in the Alpha Cen B data almost certainly arises from the window function (time sampling) of the original data. We show that when stellar activity signals are removed from the RV variations, other significant peaks in the power spectrum of the window function are coincidentally suppressed, leaving behind a spurious yet apparently-significant `ghost' of a signal that was present in the window function's power spectrum \emph{ab initio}. Even when fitting synthetic data with time sampling identical to the original data, but devoid of any genuine periodicities close to that of the planet candidate, the original model used to infer the presence of Alpha Cen Bb leads to identical conclusions: viz., the $3\sigma$ detection of a half-a-metre-per-second signal with $3.236$~d period. Our analysis underscores the difficulty of detecting weak planetary signals in RV data, and the importance of understanding in detail how every component of an RV data set, including its time sampling, influences final statistical inference. 
\end{abstract}
%------------------------------------------------------------------------------------------------------------------------------
%------------------------------------------------------------------------------------------------------------------------------
\section{Introduction}\label{sec:intro}
%------------------------------------------------------------------------------------------------------------------------------

The announcement by \citet{dumusque2012nature} of the detection of a planet around the modestly-active, \mbox{K1 V} star \aCB~generated exceptional excitement in the astronomy community: if verified, the claimed planet \aCB b would be the closest exoplanet to Earth ever discovered (distance: $1.34$~pc), and the lowest-minimum-mass planet detected around a Solar-type star. The claimed planet has an orbital period of $3.2357\pm0.0008$~d, and an estimated minimum mass of $1.13\pm0.09$~$M_\oplus$, inferred from an RV semi-amplitude of $0.51\pm0.04$~\mps.

\citeauthor{dumusque2012nature} (hereafter D12) obtained $459$ HARPS RV datapoints, along with ancillary line width (FWHM), bisector inverse slope (\bis) and chromospheric activity (\lrhk) time series, over a period of 4 years. %The D12 RVs are dominated by a long-term linear trend, due to the orbit of \aCB\ around the centre of mass of the $\alpha\,{\rm Cen}$ binary system. Once this trend is subtracted, a gradual rise and fall over the 4-year span of the observations is evident, as well as variability on shorter time-scales.
D12 used a variety of mathematical transformations to try to filter out many sources of RV variance, including starspots, photospheric granulation, and binary motion due to the presence of companion star \aCA. The amplitude of the binary-motion signal they removed was on the order of hundreds of \mps; the combined amplitude of nuisance signals ascribed to stellar activity was on the order of a few~\mps; and the final, putative planetary signal they isolated had a semi-amplitude of $\sim0.5$~\mps. For comparison, HARPS' long-term precision is $0.8$~\mps\ \citep{mayor2003}.

The detection, however, was a contentious one, with some voicing reservations about the modelling approach used to infer the planet's presence  \citep[e.g.,][]{hatzes2012,hatzes2013}. For example, short-term rotational activity was modelled using sinusoidal waves at the rotational period of the star, and a varying number of harmonics -- essentially, a truncated Fourier series. \textcolor{black}{This is not necessarily a good approximation in the case of an \emph{evolving} rotation signal \citep[e.g.,][]{lanza2001}, and subtracting such a model from the \drv~time series opens the possibility of inadvertently introducing periodic signals, or harmonics thereof, into the power spectrum.} Second, available activity-sensitive time series (BIS and $\log{R'_\textrm{HK}}$) were not jointly modelled with the $\Delta$RV time series, even though the rotational signal was strongest in the \lrhk\ time series. Instead, a low-pass smoothing filter was applied to the $\log{R'_\textrm{HK}}$ time series; this smoothed signal was noted to look similar to the $\Delta$RV time series, and then used to mitigate the effects of long-term activity in the \drv\ time series. Third, each of the four observing seasons was fitted independently of the others, possibly discarding information about long-term correlations in the time series, and resulting in a more complex model than would likely have been the case if fitting were performed for all seasons simultaneously. The model used to fit the $\Delta$RV time series alone, without a planet, contained $23$~free parameters, which raises the question of possible over-fitting. %In a similar vein, no model comparison was performed to quantify the evidence for the existence of the planet, i.e.\ the extent to which the data did or didn't favour a planetary model. 
These potential reservations aside, however, the signal D12 extracted certainly did seem to bear many of the hallmarks of a real planetary signal, including a coherent phase over four observing seasons, and low false alarm probability.

We previously attempted \citep[][hereafter R15]{Rajpaul2015} to address some of the aforementioned reservations by applying to the D12 data set a Gaussian process (GP) framework that we developed to model RV time series jointly with ancillary, activity-sensitive time series, allowing the activity component of an RV signal to be disentangled from dynamical components. Our modelling framework uses GP draws and derivatives thereof as basis functions to model available time series in a flexible, data-driven way, rather than e.g.\ sinusoids or other simplistic parametric models. Moreover, the entire framework is accommodated very naturally within the broader framework of Bayesian inference, allowing uncertainties to be handled in a principled way: overly-complex models are automatically penalized, nuisance parameters can be marginalized over, and rigorous model comparisons can be performed. Using this framework, we showed that the observed RV variations could be satisfactorily modelled as stellar activity alone, without requiring a planet. Yet we did not claim that \aCB b does \emph{not} exist, in part because that analysis did not provide any definitive explanation for the origin of the $3.24$~d signal uncovered by D12.

%We present here a re-analysis of D12's data, and argue that the $3.24$~d signal almost certainly arises from the window function (time sampling) of the original data.

% The balance of our paper is structured as follows. The next section (Section \ref{sec:apparent}) considers some of the properties of the $3.24$~d signal that make it consistent with a real planetary signal, as well as some warning flags that suggest the possibility of the $3.24$~d signal being pathological. Section \ref{sec:window} focuses on the window function of D12's dataset, which has relevance to Section \ref{sec:origin}, in which we use D12's original model to demonstrate the non-planetary origins of the $3.24$~d signal. Finally, Section \ref{sec:discuss} provides recommendations for future attempts at detecting low-mass exoplanets in RV data sets, and concludes.

%------------------------------------------------------------------------------------------------------------------------------
\section{An apparent planetary signal}\label{sec:apparent}
%------------------------------------------------------------------------------------------------------------------------------
To probe the origins of the planetary signal claimed by D12, we implemented exactly the planet-free model published by D12, and used it to fit the original observations. We optimised model parameters using multiple runs of a Nelder-Mead algorithm, and obtained marginally better goodness-of-fit statistics than those published by D12 (likely because D12 approximated the optimisation problem as a linear one; X.\ Dumusque, \emph{pers. comm.,} 2015). Nevertheless, our RV residuals corresponded almost exactly to those published by D12, and the $3.24$~d signal appeared, with the expected $>3\sigma$ significance.

Studying the power spectra of the \emph{periodic} components of D12's model (see Section \ref{sec:origin}), in combination with varying levels of noise, proved inconclusive; none of these components appeared to give rise to $3.24$~d periodicities. This was consistent with D12's own bootstrap tests to ascertain whether the $3.24$~d signal was introduced during the process of eliminating stellar signals. We next ran a Bayesian model comparison test of D12's baseline model vs.\ D12's planet model. We placed uniform priors on all parameters, with prior domains set arbitrarily to ten times the $\pm1\sigma$ confidence intervals published by D12; model evidences ($Z$) were computed using the \textsc{MultiNest} nested-sampling algorithm \citep{multinest2008,multinest2009,multinest2013}. The planet model was favoured very strongly over the simpler baseline model, with a Bayes factor on the order of $10^5$. Moreover, $\log{Z}$ for the planet model increased approximately linearly when artificially increasing the number of datapoints used in the modelling (up to a maximum of $459$).

However, some further tests suggested -- without proving outright that \aCB b does not exist -- that a planetary origin for the $3.24$~d signal is unlikely. These tests are sketched below.

First, we injected into D12's RV observations different $0.51$~\mps\ sinusoidal signals, with random phases, and periods uniformly distributed between $3.24\pm2.0$~d, excluding $3.24\pm0.2$~d. We considered $100,000$ such signals. In $87\%$ of cases, the normalised least-squares/Lomb-Scargle (LS) power at the period of the injected signal was, after first subtracting the dominant binary signal from the data, higher than the power of the strongest signals between $3.24\pm0.02$~d. (For a given period $P$, the LS periodogram of a signal $f(t)$ represents a test statistic for the corresponding problem of testing between two hypotheses: viz.\ that $f(t)$ represents white Gaussian noise, vs.\ a coherent sinusoidal signal with period $P$, based on a fit using a least-squares metric.)

%------------------------------------------
% Figure: schematic GP fitting scheme
%------------------------------------------
\begin{figure}
\begin{center}
\includegraphics[width=\columnwidth]{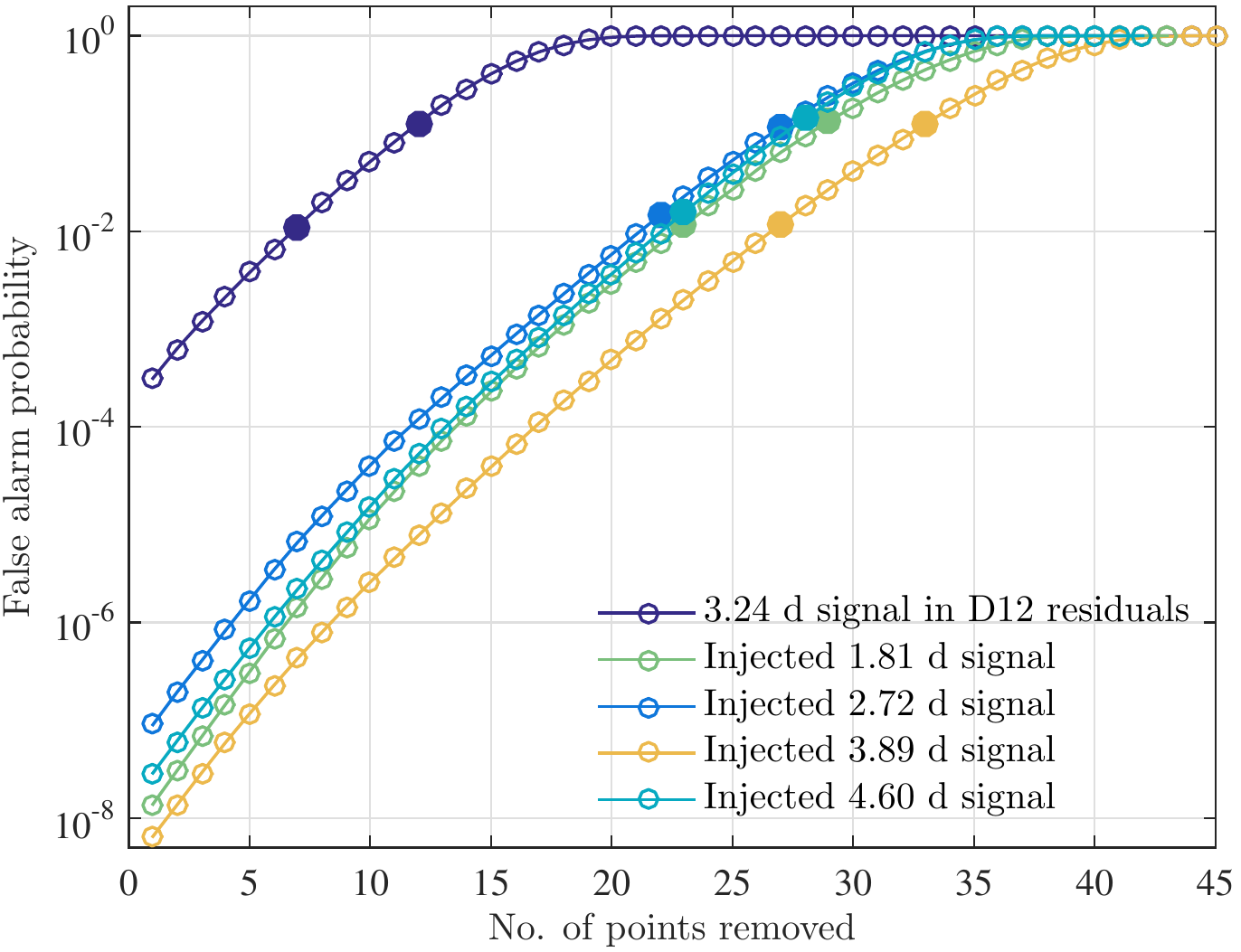}
\caption[]{Illustration of a few $0.51$~\mps\ signals, injected into D12's published residuals, which are more robust to removal of data points than D12's signal. Solid markers correspond to points where signals first exceed $1\%$ and $10\%$ false alarm probability (FAP) thresholds.}
\label{fig:remove1}
\end{center}
\end{figure}
%------------------------------------------
%------------------------------------------
% Figure: schematic GP fitting scheme
%------------------------------------------
\begin{figure}
\begin{center}
\includegraphics[width=\columnwidth]{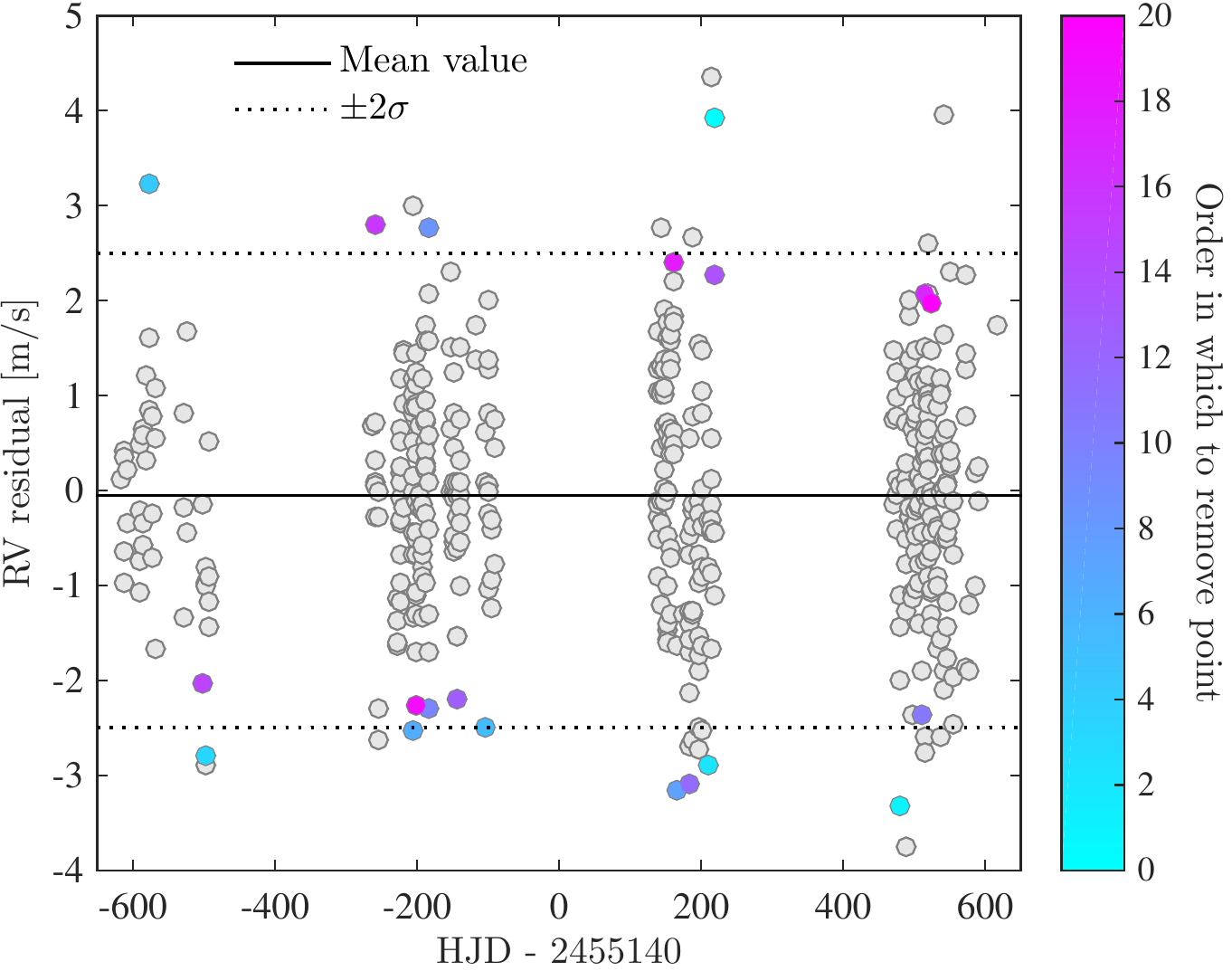}
\caption[]{Illustration of the points which, when removed from D12's observations, `destroy' the $3.24$~d signal in the published residuals. The colour-coded order in which points are to be removed corresponds to Fig.\ \ref{fig:remove1}; e.g.\ removing points $1$ to $12$ will suffice to boost the signal's FAP above $10\%$.}
\label{fig:remove2}
\end{center}
\end{figure}
%------------------------------------------

Second, we considered $100,000$ sets of random Gaussian noise, drawn at times identical to D12's observations, and with standard deviations on each point corresponding to D12's HARPS error estimates. We found that $18\%$ of these noise signals contained more normalised LS power specifically at $3.24\pm0.02$~d than was present in D12's binary-subtracted observations. Moreover, \emph{all} of these time series of noise contained at least one signal between $2$~d and $10$~d with lower false alarm probabilities (FAPs) than any signals between $3.24\pm0.02$~d in D12's binary-subtracted data.

Third, we found that the statistical significance of D12's $3.24$~d signal could be destroyed by removing as few as $12$ (non-random) points from the residuals. These points were distributed through all four observing seasons, and happened to correspond to unusually large amplitude residuals ($>2.5~\mps$). On the other hand, it was a straightforward exercise to find injected signals with amplitude and period similar to D12's signal, but which were twice as robust to data points being removed. See Fig.\ \ref{fig:remove1} and Fig.\ \ref{fig:remove2}\textcolor{black}{; cf.\ \citet{hatzes2013}.}

Fitting the original data using the framework described in R15 proved puzzling. Our best-fitting GP models had no $3.24$~d signal in the residuals; curiously, however, we found that the presence of the $3.24$~d signal seemed to be sensitive to how we fitted the long-term binary signal in the data (e.g.\ subtracting it before fitting the rest of the model vs.\ fitting everything simultaneously), as well as to whether we binned observations to one point per night (by default, we did \emph{not} bin any data). This motivated us to look beyond our activity models -- and in particular, to make a closer study of the window function for D12's observations.

%------------------------------------------------------------------------------------------------------------------------------
\section{A closer look at the window function}\label{sec:window}
%------------------------------------------------------------------------------------------------------------------------------

If a signal is sampled at discrete times, the so-called window function (i.e.\ vector of discrete sampling times) imprints periodicities on the signal, and the LS periodogram of the window function gives us information about the significance of these periodicities associated with sampling. At first glance, the LS periodogram of the window function for D12's data set is devoid of interesting features. It is dominated by a tight cluster of peaks around $1$~d, arising due to periodicities in the observing times imposed by the Solar and sidereal day. The strongest feature (at $0.9978$~d) has an LS FAP tens of thousands of times lower than any other peak at periods shorter than a few months.

\begin{figure*}
\begin{center}
\includegraphics[width=\textwidth]{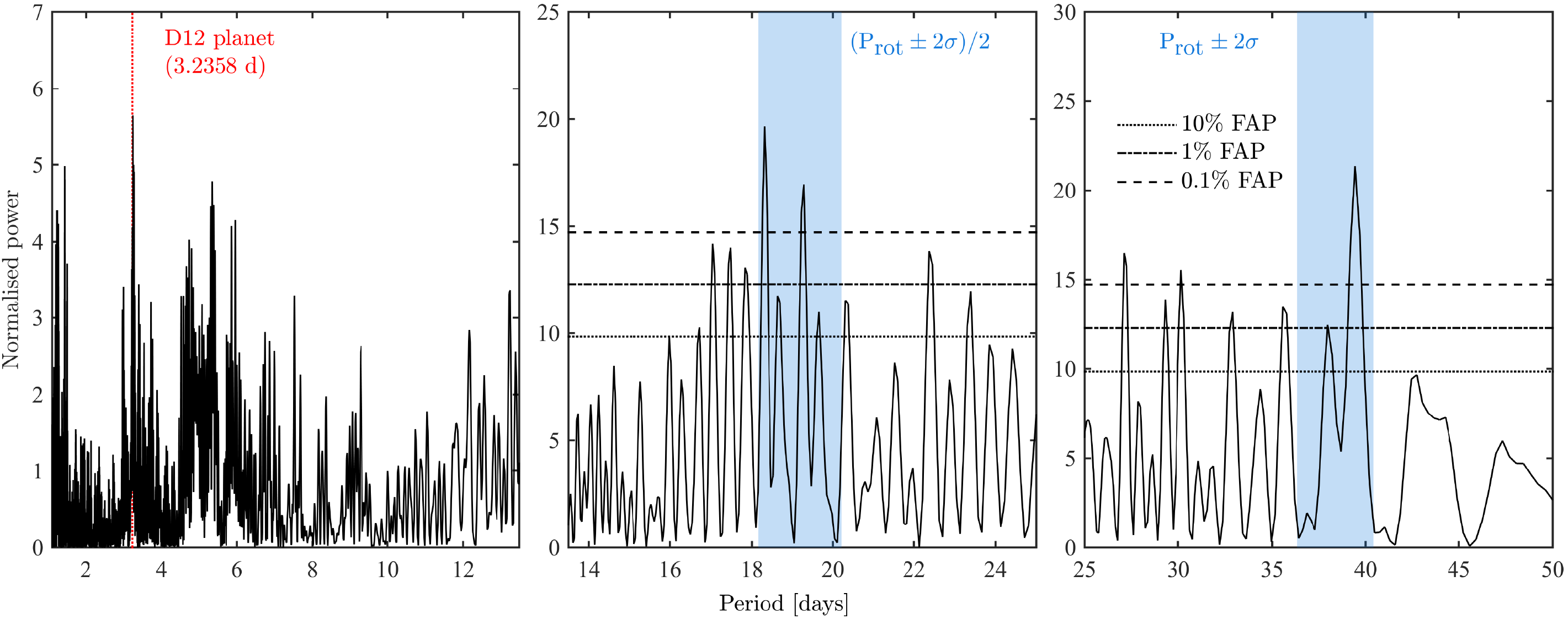}
\caption[]{LS power spectrum of the window function of D12's observations. The $\argmax$ of the power spectrum on three non-trivial domains corresponds to the period of D12's planetary candidate (left panel), the estimated rotation period of \aCB\ (right), and the first harmonic of the rotation period (middle).}
\label{fig:window_func}
\end{center}
\end{figure*}

However, when masking off this `over-saturated' region of the power spectrum, the following become apparent. First, the strongest feature between (just over) $1$~d and approximately two weeks is a complex of peaks around $3.249$~d, including a peak at $3.236$~d. Second, on time-scales between (just over) $1$~d and approximately three months, the three strongest peaks in the power spectrum are consistent with $\textrm{P}_\textrm{rot}\pm2\sigma$ or $\textrm{P}_\textrm{rot}/2\pm\sigma$, where $\textrm{P}_\textrm{rot}$ is a stellar rotation period for \aCB~derived by D12, and $\sigma$ is the uncertainty on the period (D12 derived slightly different rotation periods for different observing seasons, which they attributed to differential rotation of \aCB). See Fig.\ \ref{fig:window_func}.

%------------------------------------------
% Figure: schematic GP fitting scheme
%------------------------------------------
%------------------------------------------

\begin{figure}
\begin{center}
\includegraphics[width=\columnwidth]{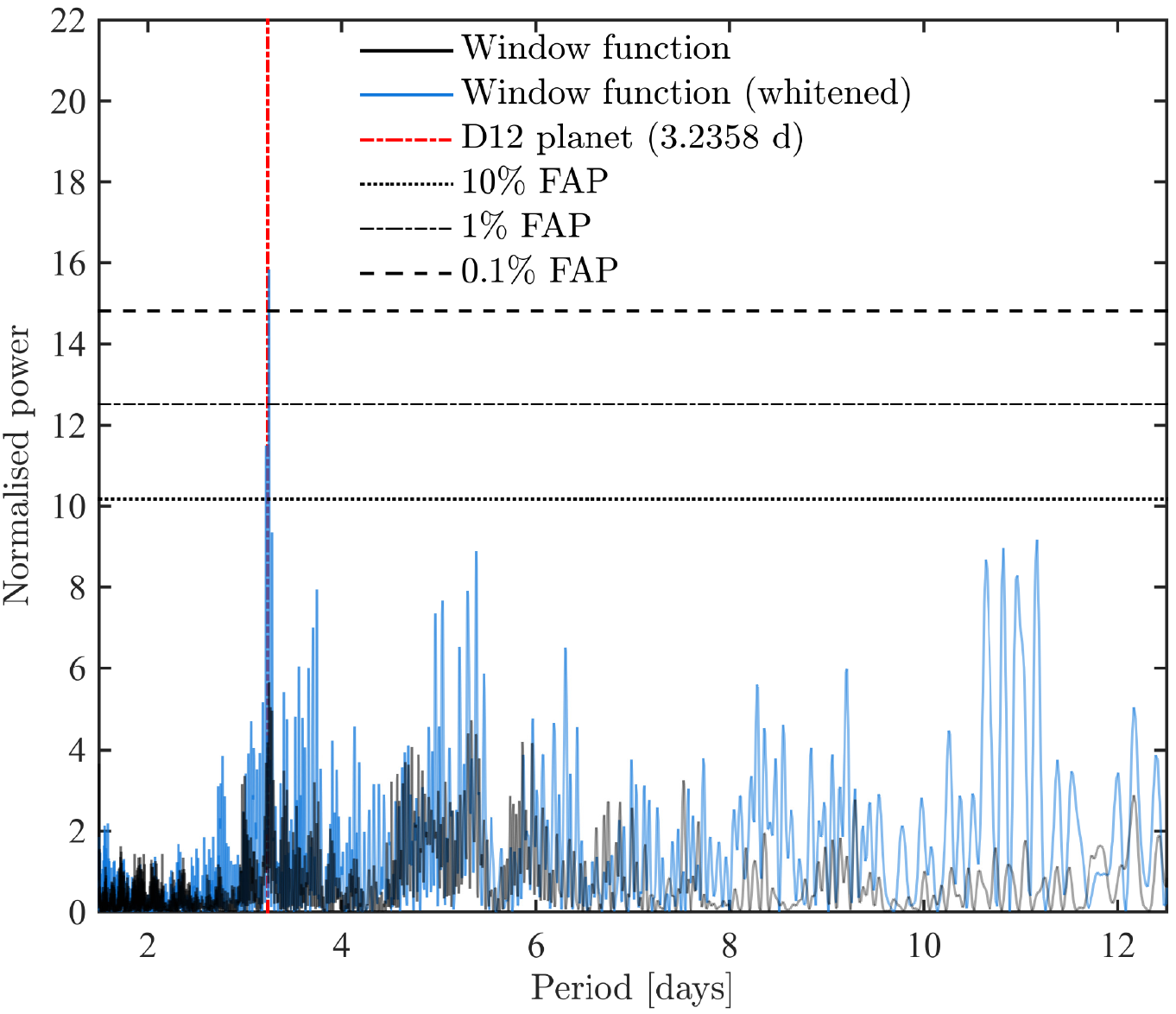}
\caption[]{Power spectrum of the `whitened' window function, after removing four fitted sinusoids with periods constrained to lie between $\textrm{P}_\textrm{rot}\pm2\sigma$ or $\textrm{P}_\textrm{rot}/2\pm\sigma$ (ranges taken from D12's best-fitting model), and unconstrained amplitudes and phases. For comparison, D12's activity model included eight different sinusoidal components.}
\label{fig:whitened}
\end{center}
\end{figure}

It seems a remarkable coincidence that the peaks around $3.249$~d correspond (within $20$ minutes) to the period of D12's planet candidate, \emph{and} that they happen to be the most significant peaks in a sizeable region of the power spectrum. Ordinarily, the formal FAPs ($>99\%$) of these peaks might lead to them being ignored. However, it is a further coincidence that the power spectrum contains significant structure at periods consistent with the stellar rotation period, $\textrm{P}_\textrm{rot}$, and its first harmonic. By `whitening' the power spectrum and removing these periodicities (e.g.\ by fitting and removing sinusoids at these periods, exactly as was done with D12's model), the statistical significance of the peaks at $3.24$~d can be enhanced. Indeed, as Fig.\ \ref{fig:whitened} confirms, the $\sim3.24$~d signals can readily be boosted above a $3\sigma$ significance threshold simply by suppressing `rotational activity' signals in the window function. Of course, the window function contains no real rotational activity signals, except as an unfortunate numerical coincidence. 

To probe further the extent to which the window function gives rise to $\sim3.24$~d periodicities, we computed the LS power spectra for $10,000$ polynomials with degree $<10$, and uniformly-random coefficients, all with time sampling identical to D12's observations. In \emph{every} case, the strongest periodicity between about $1.1$~d and $13.5$~d was found at $3.249$~d. When fitting and subtracting a polynomial from one of these randomly-generated polynomial `observations,' the power spectrum was flattened out (as expected, since exact fits were possible). However, when first adding non-white noise to the polynomials (e.g.\ random walk noise, periodic and quasi-periodic signals, GP draws, or combinations thereof) and then fitting and subtracting a polynomial, we observed different behaviour. 

In cases where the injected signals `drowned out' the polynomial signals (i.e.\ the window function imprinted on a polynomial), the power spectra of the residuals tended to match closely those of the injected signals. However, in other cases where the injected signals' contribution to the total signal variance was much smaller than the polynomial contribution, the original polynomials could be fit well, but \emph{not} perfectly. This meant that the residuals still contained a small signal bearing the imprint of the window function, so very often in these cases the most significant period (usually between $1.1$~d and $13.5$~d) in the residuals turned out to be $3.249$~d, or a period very close to (within $1\%$ of) $3.249$~d. This was not surprising, as our earlier test confirmed that virtually \emph{any} polynomial with the same sampling as D12's observations would have in its power spectrum a $3.249$~d signal as a local maximum. Slight ($<1\%$) deviations from $3.249$~d in the latter tests could be attributed to biases in the power spectrum introduced by the injected signals. Furthermore, when suppressing other, more significant components of the window function (as before; see Fig.\ \ref{fig:whitened}), the significance of the $\sim3.24$~d signals could be boosted.

%------------------------------------------------------------------------------------------------------------------------------
\section{The true origin of the planetary signal}\label{sec:origin}
%------------------------------------------------------------------------------------------------------------------------------
D12's baseline model used to fit the RVs contained three components, which modelled, respectively: (1) the binary signal; (2) long-term magnetic activity; and (3) stellar rotational activity. Component (1) comprised a polynomial, second order (quadratic) in time, with the same three coefficients used for all observing seasons. Component (2) was proportional to the low-pass filtered \lrhk\ series; the only free parameter here was a scaling factor to convert the \lrhk\ to \drv~variations. Component (3) included sinusoidal terms, different for each season; in total, it contained $19$ free parameters. D12 also considered a model with an added sinusoid term, (4), to describe a possible planetary contribution to the \drv\ time series. This took the form $\Delta {\text{R}}{{\text{V}}_{{\text{planet}}}} = {K_1}\sin (2\pi t/P) + {K_2}\cos (2\pi t/P)$; equivalently, $\Delta {\text{R}}{{\text{V}}_{{\text{planet}}}} = K\sin (2\pi t/P + \phi )$. All components of the model were fitted simultaneously.

When computing the LS power spectrum of component (1) of the best-fitting model, i.e.\ the polynomial component, the normalized LS power at $3.25\pm0.1$~d was found to be $99.3\%$ of the normalized power from the raw~\drv\ observations (which included a binary signal, long-term magnetic activity variations, rotational activity, and unknown contributions). In fact, we had to inject into the raw~\drv\ observations sinusoids with semi-amplitudes on the order of tens of \mps\ (assuming periods on the order of a few days) to obtain comparable LS power to that observed at $3.25\pm0.1$~d. 

This suggests the following scenario: a signal at $\sim3.24$~d was present in the original observations; however, it arose not from a planet, but from the window function. As with the tests described at the end of Section \ref{sec:window}, D12's imperfect model meant that the imprint or `ghost' of the window function remained in the model residuals. Moreover, because other significant components of the window function happen to be suppressed when removing stellar activity, i.e.\ through model component (3), the $\sim3.24$~d signals end up being \emph{boosted} in the window function's imprint on the residuals.

To test this scenario, we generated synthetic observations as follows. Components (1) and (2) of D12's original model, with free parameters set at their maximum \emph{a posteriori} (MAP) values, were taken to represent (a) binary, and (b) long-term magnetic variations. We then used the framework from R15 to train a GP with quasi-periodic covariance kernel on the `observed' rotational activity (\drv~observations minus binary and magnetic activity models), in tandem with \lrhk~and~ \bis~observations. We took the posterior mean of the GP for the \drv~series to represent rotational activity, (c). Finally, components (a), (b), and (c) were co-added with white Gaussian noise to form a set of synthetic observations.

We avoided using component (3) of D12's model to synthesise the rotational activity: otherwise the synthetic observations would have been generated entirely by D12's original model, and it would have been na\"ive to expect anything but an excellent fit, with white, Gaussian residuals (see Section \ref{sec:window}). Instead, using a GP allowed us to synthesise observations with the same covariance properties -- (quasi)periodicities, evolution time-scales, smoothness, noise levels, etc.\ -- as the observed rotational activity, regardless of the extent to which D12's simple harmonic model might've failed to describe such variations. Furthermore, the GP framework meant that any periodicities not also present in the ancillary time series (i.e.~\lrhk, \bis) would \emph{not} be captured by the \drv\ model. We confirmed that the power spectrum of component (c) of our model was free of any significant periodicities on time-scales of a few days; and it certainly contained no signals near $3.24$~d.

\begin{figure}
\begin{center}
\includegraphics[width=\columnwidth]{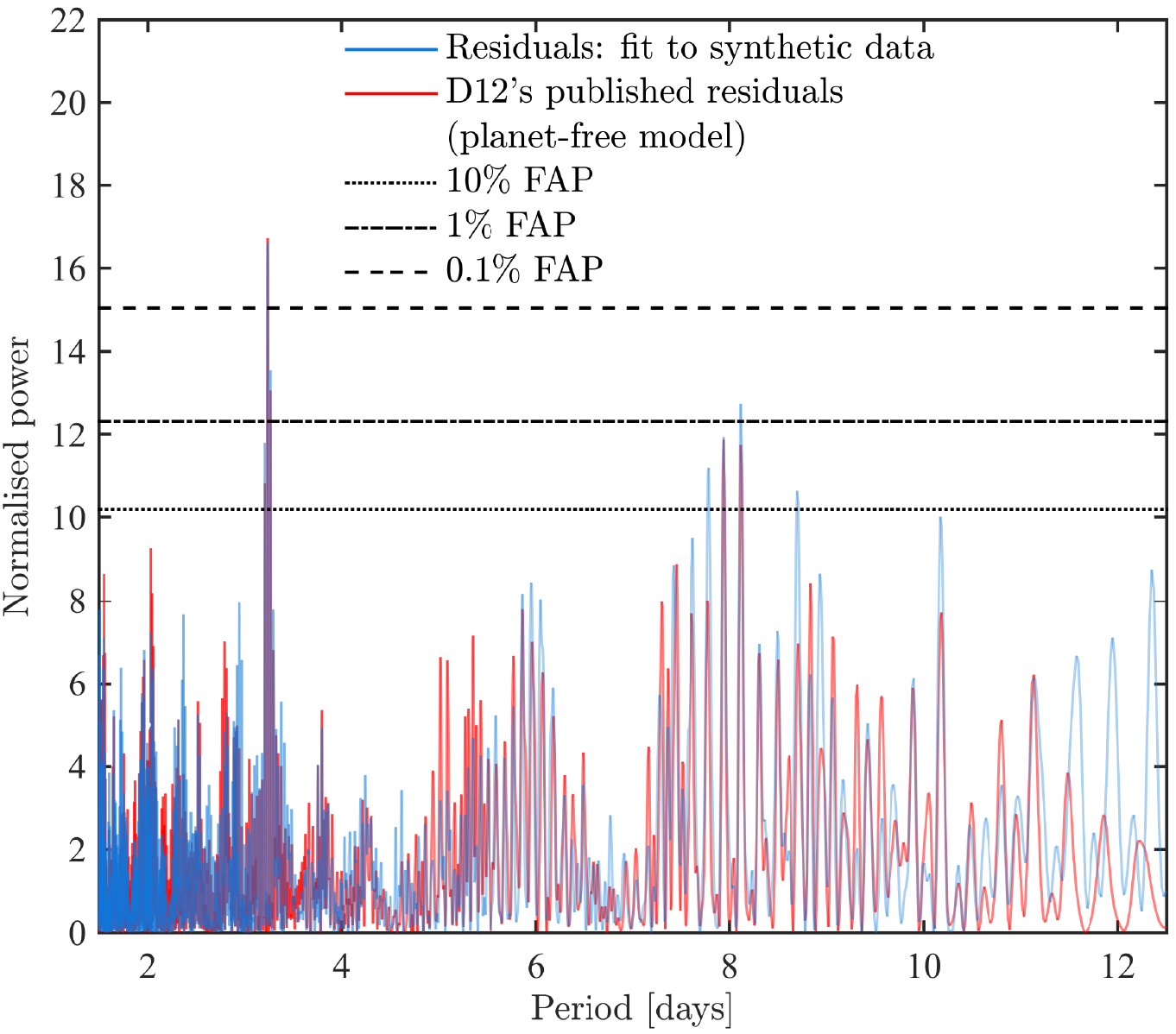}
\caption[]{Power spectrum of the residuals from synthetic, planet-free data (with realistic covariance properties, and time sampling identical to D12's data), obtained from a MAP fit with D12's original model. In spite of all, a strong signal appears at $3.2358$~d, corresponding to D12's planet candidate.}
\label{fig:GP_residuals}
\end{center}
\end{figure}

We next used D12's model to fit our baseline set of synthetic observations -- to be quite clear, a set of observations synthesized only from non-planetary components of D12's model, a conservative activity model for the original observations, and white noise. As before, uninformative priors were placed on all parameters in D12's model, and  parameter inference was performed using \textsc{MultiNest}. The strongest feature in the power spectrum of the residuals from the MAP fit was a significant peak at $\sim3.236$~d, with estimated FAP of $0.04\%$; see Fig.\ \ref{fig:GP_residuals}. When adding a zero-eccentricity Keplerian (sinusoidal) component to D12's overall model, we obtained an RV semi-amplitude for the Keplerian of $K=0.46\pm0.03$~\mps, and a period of $3.2358\pm0.0002$~d. \textcolor{black}{As before, the more complex sinusoidal model was strongly favoured over the baseline model, with a Bayes factor of $\sim10^5$.}

We repeated these tests 10 times using different `observations', synthesized by perturbing the MAP parameters from components (1) and (2) of D12's model, within their $1\sigma$ confidence intervals; by using different GP draws for the rotational activity model, within a $1\sigma$ envelope of the GP posterior mean function; and by using different seeds when generating additive white Gaussian noise. In each case, a significant signal appeared at $\sim3.236$~d in the residuals, and when including a Keplerian in the model, a MAP period consistent with $3.2358$ (within $1\sigma$) was obtained. However, a range of semi-amplitudes was obtained -- the smallest was $K=0.33\pm0.03$~\mps, and the largest, $K=0.53\pm0.04$~\mps. For comparison, the planet claimed by D12 had best-fitting parameters $K=0.51\pm0.04$~\mps, and $P=3.2357\pm0.0008$~d.

It is exceptionally implausible that synthetic data simply with the same time sampling and activity-related covariance properties as D12's dataset, but no planet, would consistently lead to the detection of the same planet claimed by D12, if indeed D12's planet were real. We thus conclude that the $3.24$~d signal that D12 originally recovered was most likely nothing more than a `ghost' of the window function, amplified during the modelling process to non-trivial significance levels.
%------------------------------------------------------------------------------------------------------------------------------
\section{Discussion and conclusions}\label{sec:discuss}
%------------------------------------------------------------------------------------------------------------------------------
%------------------------------------------
% Figure: schematic GP fitting scheme
%------------------------------------------
%------------------------------------------

We have demonstrated that the $3.24$~d signal observed in the \aCB~data almost certainly arises from the window function of the original data. When stellar activity signals are filtered out from the RV variations, other significant peaks in the power spectrum of window function are coincidentally suppressed, leaving behind a spurious yet apparently-significant `ghost' of a 3.24~d signal that was present in the window function's power spectrum \emph{ab initio}.

% We showed that even when fitting synthetic data that have the same time sampling as the original data, but that are by design devoid of any genuine periodicities close to the period of the planet candidate, the original model used to infer the presence of Alpha Cen Bb leads to identical conclusions: viz., the $3\sigma$ detection of a half-a-metre-per-second signal with $3.236$~d period.

Our analysis underscores the difficulty of detecting weak planetary signals in RV data, and the importance of understanding in detail how every component of an RV data set, including its time sampling, influences final statistical inference. In future attempts at detecting low-mass exoplanets in RV data sets, it would be prudent to study the window function of the observations carefully. A planet candidate should be treated with extreme circumspection if its period happens to coincide closely with an $\argmax$ of the window function's power spectrum on some non-trivial domain; in the case of the D12 data, $3.24$~d happened to be the $\argmax$ of the window function's power spectrum on an interval between about $1$~d and two weeks. Additionally, D12's data set was particularly pathological because the window function happened to contain periodicities that coincided with the stellar rotation period of \aCB, and its first harmonic; when these signals were filtered out, the significance of the $3.24$~d signal was preferentially boosted.

\textcolor{black}{We alluded to a number of other tests we believe worth carrying out when considering the reliability of planet detections from noisy, discretely-sampled signals. These include using the same model used to detect the planet instead to fit synthetic, planet-free data (with realistic covariance properties, and time sampling identical to the real data), and checking whether the `planet' is still detected; comparing the strength of the planetary signal with similar Keplerian signals injected into the original observations; performing Bayesian model comparisons between planet and no-planet models; and checking how robust the planetary signal is to datapoints being removed from the observations. As this study has demonstrated, simply considering the false alarm probability of a signal in a model's residuals, or (somewhat equivalently) checking whether a signal has a coherent phase over a full set of observations, is clearly insufficient. Even rigorous, Bayesian model comparisons do not necessarily shed light on the \emph{origins} of a planet-like signal.}

The authors propose to carry out, in the near future, a systematic study of the properties of spurious yet coherent signals that can arise in noisy datasets with real (e.g.\ HARPS-like) observing calendars. This will hopefully inform best practices for identifying or mitigating such signals; it might also prove relevant to already-published cases of planets whose existence was supported by rigorous model comparison tests, yet not by follow-up observations. Certainly, when carrying out future observations of \aCB, it would be worth ensuring that the window function of the proposed observations did not contain significant power at $3.24$~d, or $\textrm{P}_\textrm{rot}\pm2\sigma$ and harmonics thereof.

%------------------------------------------
% Figure: schematic GP fitting scheme
%------------------------------------------
%------------------------------------------
%------------------------------------------------------------------------------------------------------------------------------
\section*{Acknowledgments}
%------------------------------------------------------------------------------------------------------------------------------
The authors thank Xavier Dumusque and Andrew Collier Cameron for useful discussions. V.~R.\ thanks Merton College, the Rhodes Trust, and the National Research Foundation of South Africa for providing financial support for this work. S.~A.\ gratefully acknowledges support from the Leverhulme Trust (RPG-2012-661) and from the UK Science and Technology Facilities Council (ST/K00106X/1).
%------------------------------------------------------------------------------------------------------------------------------
\bibliography{biblio}
\bibliographystyle{mn2e}
%------------------------------------------------------------------------------------------------------------------------------

\bsp
\label{lastpage}
%------------------------------------------------------------------------------------------------------------------------------
\end{document}